\documentclass[preprint,showpacs,titlepage,aps,prd,
tightenlines,amsmath,byrevtex,nofootinbib]{revtex4}

\usepackage{graphicx}

\def\lsim{\raise0.3ex\hbox{$\;<$\kern-0.75em\raise-1.1ex
\hbox{$\sim\;$}}}
\def\gsim{\raise0.3ex\hbox{$\;>$\kern-0.75em\raise-1.1ex
\hbox{$\sim\;$}}}

\begin{document}

\preprint{hep-ph/0609286}

\title{Resolving Eight-Fold Neutrino Parameter Degeneracy by \\
Two Identical Detectors with Different Baselines
}

\author{Takaaki Kajita$^{1}$}
\email{kajita@icrr.u-tokyo.ac.jp}
\author{Hisakazu Minakata$^{2}$}
\email{minakata@phys.metro-u.ac.jp}
\author{Shoei Nakayama$^{1}$}
\email{shoei@suketto.icrr.u-tokyo.ac.jp}
\author{Hiroshi Nunokawa$^{3}$}
\email{nunokawa@fis.puc-rio.br}
\affiliation{
$^1$Research Center for Cosmic Neutrinos, Institute for Cosmic Ray Research, University of Tokyo, Kashiwa, Chiba 277-8582, Japan \\
$^2$Department of Physics, Tokyo Metropolitan University, Hachioji, Tokyo 192-0397, Japan \\
$^3$Departamento de F\'{\i}sica, Pontif{\'\i}cia Universidade Cat{\'o}lica 
do Rio de Janeiro, C. P. 38071, 22452-970, Rio de Janeiro, Brazil
}

\date{September 27, 2006}

\vglue 1.4cm

\begin{abstract}

We have shown in a previous paper  that two identical detectors 
with each fiducial mass of 0.27 megaton water, 
one in Kamioka and the other in Korea, which receive the (anti-) muon 
neutrino beam of 4 MW power from J-PARC facility 
have potential of determining the neutrino mass hierarchy and discovering 
CP violation by resolving the degeneracies associated with them. 
In this paper, we point out that the same setting has capability 
of resolving the $\theta_{23}$ octant degeneracy in region where  
$\sin^2 2\theta_{23} \lsim 0.97$ at 2 standard deviation 
confidence level even for very small values of $\theta_{13}$. 
Altogether, it is demonstrated that one can solve all the eight-fold neutrino 
parameter degeneracies {\em in situ} by using the 
Tokai-to-Kamioka-Korea setting if $\theta_{13}$ 
is within reach by the next generation superbeam experiments. 
We also prove the property called ``decoupling between the degeneracies'', 
which is valid to first order in perturbation theory of the earth matter 
effect,  that guarantees approximate independence between analyses to 
solve any one of the three different type of degeneracies.

\end{abstract}


\maketitle


\section{Introduction}

Physics of neutrinos has entered into a new stage 
after establishment of the mass-induced neutrino oscillation due to 
the atmospheric~\cite{atm_evidence}, the accelerator 
~\cite{K2K_evidence, MINOS}, and 
the reactor neutrino~\cite{KL_evidence} experiments, 
confirming the earlier discovery~\cite{SKatm,solar,KamLAND} 
and identifying the nature of the phenomenon. 
In the new era, the experimental endeavors will be focused on search for 
the unknowns in neutrino masses and the lepton flavor mixing, 
$\theta_{13}$, the neutrino mass hierarchy, and CP violation. 
On the theory side, various approaches toward understanding physics of 
lepton mixing and the quark-lepton relations are extensively pursuit 
\cite{moha-smi}, which then further motivate 
precision measurement of the lepton mixing parameters. 
We will use the standard notation~\cite{PDG} of the lepton mixing matrix, 
the Maki-Nakagawa-Sakata (MNS) matrix~\cite{MNS}, 
throughout this paper.

It was recognized sometime ago that there exists problem of 
parameter degeneracy 
which would act as an obstacle against precision measurement of 
the lepton mixing parameters. 
The nature of the degeneracy can be understood as 
the intrinsic degeneracy~\cite{intrinsic}, 
which is duplicated by the unknown sign of atmospheric 
$\Delta m^2$~\cite{MNjhep01} (hereafter, ``sign-$\Delta m^2$ degeneracy'' for 
simplicity) 
and by the possible octant ambiguity of $\theta_{23}$~\cite{octant} 
that exists if $\theta_{23}$ is not maximal. 
For an overview of the resultant eight-fold degeneracy, see e.g., 
\cite{BMW,MNP2}.

In a previous paper~\cite{T2KK}, we have shown that 
the identical two detector setting in Kamioka and in Korea with 
each fiducial mass of 0.27 Mton water, which 
receives the identical neutrino beam from the J-PARC facility 
can be sensitive to the neutrino mass hierarchy and CP violation 
in a wide range of the lepton mixing parameters, $\theta_{13}$ 
and the CP phase $\delta$.
It is the purpose of this paper to point out that the same setting 
has capability of resolving the $\theta_{23}$ octant degeneracy 
to a value of $\theta_{23}$ which is rather close to the maximal, 
$\sin^2 2 \theta_{23} < 0.97 (0.94)$ at 2 (3) standard deviation 
confidence level (CL). 
It is achieved by detecting solar-$\Delta m^2$ scale oscillation effect in 
the Korean detector. 
Together with the sensitivities to resolution of the degeneracy 
related to the mass hierarchy and the CP phase discussed in 
the previous paper, we demonstrate that 
the Kamioka-Korea two detector setting is capable of solving 
the total eight-fold parameter degeneracy.
We stress that resolving the degeneracy is crucial to precision 
measurement of the lepton mixing parameters on which we make 
further comments at appropriate points in the subsequent discussions. 
We also emphasize that it is highly nontrivial that one can formulate 
such a global strategy for resolving all the known degeneracies 
(though in a limited range of the mixing parameters) 
only with the experimental apparatus using conventional muon neutrino 
superbeam.\footnote{
It may be contrasted to the method for resolving the degeneracy 
based on neutrino factory examined in~\cite{donini}; 
It uses a 40 kton magnetized iron calorimeter and a 4 kton emulsion chamber, 
and conventional $\nu_{\mu}$ beam watched by 
a 400 kton water Cherenkov detector. 
}

In some of the previous analyses including ours~\cite{T2KK,resolve23}, 
people often tried to resolve the degeneracy of a particular type 
without knowing (or addressing) the solutions of the other types of 
degeneracies. 
But, then, the question of consistency of the procedure immediately arises; 
Can one solve the degeneracy of type A without knowing the solutions 
of the other degeneracies B and C? 
Does the obtained solution remain unchanged when the assumed 
solutions for the other type of degeneracies are changed to the 
alternative ones?, etc. 
We answer to these questions in the positive in experimental settings 
where the earth matter effect can be treated as perturbation. 
We do so by showing that the resolution of the degeneracy of a 
particular type decouples from the remaining degeneracies, 
the property called the ``decoupling between the degeneracies'' 
in this paper.

In Sec.~\ref{how}, we present a pedagogical discussion of how the 
eight-fold degeneracy can be lifted by measurement with  
the Kamioka-Korea two detector setting. 
In Sec.~\ref{decoupling}, we prove the ``decoupling'' and make 
a brief comment on its significance.
In Sec.~\ref{23degeneracy}, we discuss some characteristic features 
of the $\nu_{e}$ and $\bar{\nu}_{e}$ appearance probabilities 
that allow the Kamioka-Korea identical two detector setting to 
resolve the $\theta_{23}$ octant degeneracy. 
In Sec.~\ref{sensitivity}, the actual analysis procedure 
and the obtained sensitivities for solving the $\theta_{23}$ degeneracy 
are described in detail. 
In Sec.~\ref{revisit}, we reexamine the sensitivities to 
the mass hierarchy and CP violating phase by using our new code 
with disappearance channels and additional systematic errors.
In Sec.~\ref{summary}, we give a summary and discussions.

\section{How the identical two detector system solves the eight-fold degeneracy?}
\label{how}

We describe in this section how the eight-fold parameter degeneracy
can be resolved by using two identical detectors, one placed at a medium 
baseline distance of a few times 100 km, and the other at $\sim$1000 km or so. 
We denote them as the intermediate and the far detectors, respectively, 
in this paper. 
Whenever necessary we refer the particular setting of Kamioka-Korea 
two detector system, but most of the discussions in this and the next 
sections are valid without the specific setting.

To give the readers a level-one understanding we quote here, 
ignoring complications, 
which effect is most important for solving which degeneracy: 

\begin{itemize}

\item

The intrinsic degeneracy; 
Spectrum information solves the intrinsic degeneracy.

\item

The sign-$\Delta m^2$ degeneracy; 
Difference in the earth matter effect between the intermediate and the far detectors 
solves the sign-$\Delta m^2$ degeneracy. 

\item

The $\theta_{23}$ octant degeneracy; 
Difference in solar $\Delta m^2$ oscillation effect 
(which is proportional to $c^2_{23}$) between the intermediate 
and the far detectors solves the $\theta_{23}$ octant degeneracy. 

\end{itemize}

\begin{figure}[htbp]
\begin{center}
\includegraphics[width=0.49\textwidth]{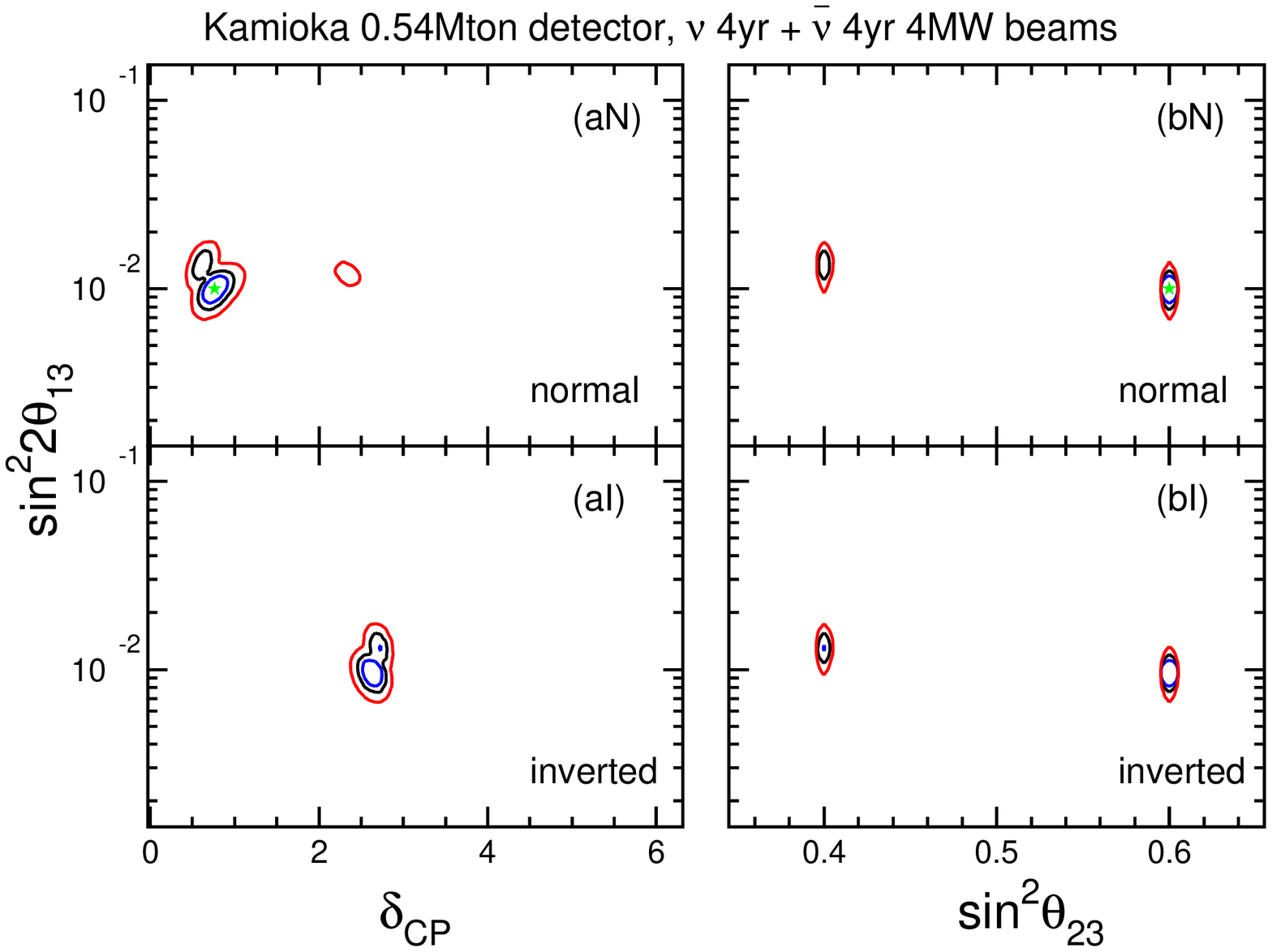}
\includegraphics[width=0.49\textwidth]{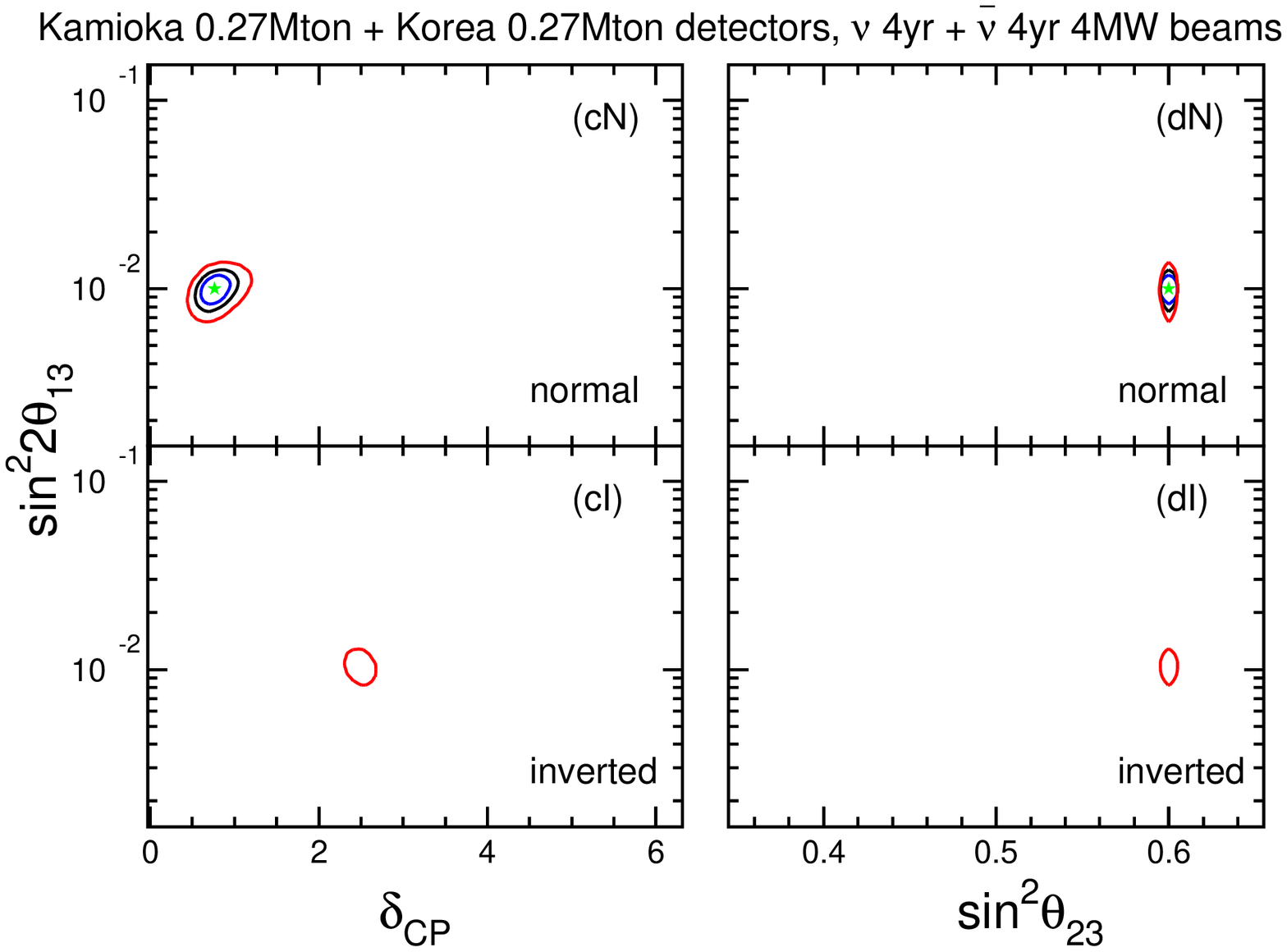}
\end{center}
\caption{The region allowed in $\delta-\sin^2 2\theta_{13}$ and 
$\sin^2 \theta_{23}-\sin^2 2\theta_{13}$ spaces 
by T2K II (left four panels) and by the 
Kamioka-Korea two detector setting (right four panels) in both of which  
4 years of neutrino plus 4 years of anti-neutrino running are assumed.  
The upper (lower) four panels show the allowed region for
the positive (negative) sign of $\Delta m^2_{31}$. 
The detector fiducial volumes of T2K II and Kamioka-Korea settings 
are assumed to be 0.54 Mton and each 0.27 Mton, respectively, 
and the beam power of J-PARC is assumed to be 4 MW. 
The baseline to the Kamioka and Korea detectors are, 
295 km and  1050 km, respectively. 
The true solution is assumed to be located at 
$\sin^2{2\theta_{13}}$=0.01, $\sin ^2 \theta_{23}$=0.60
and $\delta$=$\pi/4$  
with positive sign of $\Delta m_{31}^2 (=+2.5\times 10^{-3}$ eV$^2$), which is 
indicated by the green star. 
The solar mixing parameters are fixed as
$\Delta m_{21}^2 = 8\times 10^{-5}$ eV$^2$ and $\sin^2{\theta_{12}}$=0.31.
Three contours in each figure correspond to
the 68\% (blue line), 90\% (black line) and 99\% 
(red line) C.L. sensitivities, which are defined
as the difference of the $\chi^2$ being 2.30, 4.61 and 9.21, respectively.
}
\label{intrinsicKamKorea}
\end{figure}

To show how the eight-fold parameter degeneracy can be resolved, 
we present in Fig~\ref{intrinsicKamKorea} a comparison between 
the sensitivities achieved by the Kamioka only setting and the 
Kamioka-Korea setting by taking 
a particular set of true values of the mixing parameters which are 
quoted in caption of Fig~\ref{intrinsicKamKorea}. 
The left four panels of Fig~\ref{intrinsicKamKorea} show the 
expected allowed regions of oscillation parameters 
in the Tokai-to-Kamioka phase-II (T2K II) setting, 
while the right four panels show the allowed regions by the 
Tokai-to-Kamioka-Korea setting.
For both settings we assume 
4 years of neutrino plus 4 years of anti-neutrino 
running\footnote{
It was shown in the previous study~\cite{T2KK} that the sensitivity 
obtained with 2 years of neutrino and 6 years of anti-neutrino 
running in the T2K II setting~\cite{T2K} is very similar to that of 
4 years of neutrino and 4 years of anti-neutrino running. 
}
and the total fiducial volume is kept to be the same, 0.54 Mton.
Some more information of the experimental setting and the 
details of the analysis procedure are described in the caption of 
Fig~\ref{intrinsicKamKorea} and in Sec.~\ref{sensitivity}.

Let us first focus on the left four panels of Fig~\ref{intrinsicKamKorea}. 
In the left-most two panels labeled as (aN) and (aI), one observes 
some left-over degeneracies of the total eight-fold degeneracy; 
If we plot the result of a rate only 
analysis without spectrum information we would have seen 
8 separate (or overlapped) allowed parameter regions.  
The $\theta_{23}$ octant degeneracy remains unresolved 
as seen in panels (bN) and (bI). 
Note that the overlapping two regions in (aN) and (aI) are 
nothing but the consequence of unresolved $\theta_{23}$ degeneracy. 
The intrinsic degeneracy, horizontal pair seen in (aN), is almost 
resolved apart from 99\% CL region at the particular set of values 
of the mixing parameters indicated above. 
The corresponding pair in (aI) is missing because the 
intrinsic degeneracy is completely lifted.
Since the matter effect plays minor role in the T2K II setting it is likely 
that the spectral information is mainly responsible for lifting the intrinsic 
degeneracy. See Sec.~\ref{intrinsic} for more about it.

Here is a brief comment on the property of the intrinsic and the 
sign-$\Delta m^2$ degeneracies. 
Because the degenerate solutions of CP phase $\delta$ satisfy 
approximately the same relationship $\delta_2 = \pi - \delta_1$ 
in both the intrinsic and the sign-$\Delta m^2$ degeneracies 
\cite{intrinsic,MNjhep01} 
(see Eqs.~(\ref{intrinsicDS}) and (\ref{signdm2DS}) in Sec.~\ref{decoupling}), 
the would-be four (one missing) regions in the panels (aN) and (aI) in 
Fig.~\ref{intrinsicKamKorea} forms a cross (or X) shape, 
with crossing connection between a pair of solutions of 
the sign-$\Delta m^2$ degeneracy.

In the right four panels of Fig.~\ref{intrinsicKamKorea} it is exhibited that 
the intrinsic degeneracy as well as $\theta_{23}$ octant degeneracy 
are completely resolved by the Kamioka-Korea 
two-detector setting at the same values of the mixing parameters, 
indicating power of the two detector method~\cite{MNplb97}. 
Namely, the comparison between the spectral shapes in Kamioka 
and in Korea located at the first and nearly the second 
oscillation maxima, respectively, supersedes a single detector 
measurement in Kamioka with the same total volume 
despite much less statistics in the Korean detector. 
We will give a detailed discussion on how $\theta_{23}$ octant 
degeneracy can be resolved by the Kamioka-Korea setting in 
Sec.~\ref{23degeneracy}, and present the details of the analysis 
in Sec.~\ref{sensitivity}.

It should be noted that the sign-$\Delta m^2$ degeneracy is also 
lifted though incompletely at the particular set of values of the 
mixing parameters as indicated in the panels (cI) 
of Fig.~\ref{intrinsicKamKorea} where only the 99\% CL regions remain. 
In fact, we have shown in our previous paper that the 
Kamioka-Korea identical two detector setting is powerful in 
resolving the sign-$\Delta m^2$ degeneracy in a wide range of the 
mixing parameters~\cite{T2KK}. 
We note that resolution of the degeneracy in turn leads to an 
enhanced sensitivity to CP violation than that of T2K II setting in 
a region of relatively large $\theta_{13}$. 
See~\cite{T2KK} for comparison 
with T2K II sensitivity. 
Altogether, we verify that the identical two detector setting in 
Kamioka-Korea with neutrino beam from J-PARC solves all the 
eight-fold parameter degeneracy {\em in situ} if $\theta_{13}$ 
is within reach by the next generation superbeam experiments 
such as T2K~\cite{T2K} and NO$\nu$A~\cite{NOvA}.

\section{Decoupling between degeneracies}
\label{decoupling}

In this section we discuss the property called the 
``decoupling between degeneracies'' which arises due to the 
special setting of baselines shorter than $\sim$1000 km. 
The content of this section is somewhat independent of the main line 
of the discussion in this paper, and the readers can skip it 
to go directly to the analysis of $\theta_{23}$ octant degeneracy 
in Secs.~\ref{23degeneracy} and \ref{sensitivity}. 
Nonetheless, the property makes the structure of analysis for resolving the 
eight-fold degeneracy transparent, and therefore it may worth to report.

The problem of decoupling came to our attention via the following path. 
In most part of the previous paper~\cite{T2KK}, we have discussed 
how to solve the sign-$\Delta m^2$ degeneracy without worrying 
about the $\theta_{23}$ octant degeneracy. 
Conversely, the authors of ~\cite{resolve23} analyzed the latter 
degeneracy without resolving the former one. 
Are these correct procedure? 
The answer is yes if the analysis procedure and the results for 
the $\theta_{23}$ degeneracy is independent of which solutions 
we take for the sign-$\Delta m^2$ degeneracy, and vice versa. 
We call this property the ``decoupling between the degeneracies''.\footnote{
Here is a concrete example for which the decoupling does not work; 
In the method of comparison between 
$\nu_{\mu}$ and $\bar{\nu}_{\mu}$ disappearance measurement for 
lifting $\theta_{23}$ octant degeneracy \cite{choubey} one in fact 
determines the combined sign of 
$\cos 2\theta_{23} \times \Delta m^2_{31}$ as noticed in \cite{resolve23}, 
and hence no decoupling.
}
%
Though discussion on this point was partially given in \cite{resolve23}, 
we present here a complete discussion of the decoupling.

Under the approximation of lowest nontrivial order in matter effect, 
we prove that the decoupling holds between the above two degeneracies, 
and furthermore that it can be generalized, 
though approximately, to the relation between any two pair of 
degeneracies among the three types of degeneracies.
To our knowledge, leading order in matter perturbation theory appears to 
be the only known circumstance that the argument goes through. 
Fortunately, the approximation is valid for the setting used in this paper 
with baseline up to $\sim$1000 km, in particular in the Kamioka-Korea 
setting. 
In the following treatment we make a further approximation that the 
degenerate solutions are determined primarily by the measurement at 
the intermediate detector. 
It is a sensible approximation because the statistics is about 10 times higher 
at the intermediate detector, and its validity is explicitly verified 
in the analysis performed in \cite{T2KK}. 

\subsection{Approximate analytic treatment of the parameter degeneracy}

To make the discussion self-contained, we start from the derivation of 
the degenerate solutions by using the matter perturbation theory \cite{AKS}, 
in which the matter effect is kept to its lowest nontrivial order.
Namely, the matter effect can be ignored in leading order in 
the disappearance channel whose oscillation probability is order of unity. 
Then, the disappearance probability 
$P(\nu_{\mu} \rightarrow \nu_{\mu})$ can be given by 
the vacuum oscillation approximation with leading order in $s^2_{13}$ 
and the solar $\Delta m^2_{21}$ corrections as 
\begin{eqnarray}
1 - P(\nu_{\mu} \rightarrow \nu_{\mu})  &=& 
\left[ 
\sin^2 2\theta_{23}
+ 4 s^2_{13} s^2_{23} 
\left(2 s^2_{23} - 1 \right)
\right]
\sin^2 \left(\frac{\Delta m^2_{31}L}{4E}\right) 
\nonumber \\
& - & c_{12}^2\sin^22\theta_{23}
\left(\frac{\Delta m_{21}^2 L}{4E}\right) 
\sin \left(\frac{\Delta m_{31}^2 L}{2E}\right).
\label{Pvac_mumu2}
\end{eqnarray}
The probability for anti-neutrino channel is the same as that for neutrino 
one in this approximation. 
One can show \cite{AKS} that the other solar $\Delta m^2_{21}$ 
corrections are suppressed further by either a small 
$\sin^2{2\theta_{13}} \lsim 0.1$, 
or the Jarlskog factor 
$J \equiv c_{12} s_{12} c_{13}^2 s_{13} c_{23} s_{23} \sin{\delta} \lsim 0.04$. 
In fact, the validity of the approximation is explicitly verified in \cite{resolve23} 
where the matter effect terms are shown to be of the order of $10^{-3}$ 
even at $L=1000$ km. 
A disappearance measurement, therefore, determines $s^2_{23}$ 
to first order in $s^2_{13}$ as 
\begin{eqnarray}
(s^2_{23})^{(1)}= (s^2_{23})^{(0)} (1+ s^2_{13}),
\label{degene-sol}
\end{eqnarray}
where 
$(s^2_{23})^{(0)}$ is the solution obtained by ignoring $s^2_{13}$. 
From the first term in Eq.~(\ref{Pvac_mumu2}), 
the two solutions of $(s^2_{23})^{(0)}$ is determined as 
$(s^2_{23})^{(0)}= \frac{1}{2} \left[ 1 \pm \sqrt{1- \sin^2{2\theta_{23}}} \right]$, 
the simplest form of the $\theta_{23}$ octant degeneracy. 
For example, $(s^2_{23})^{(0)} = 0.4$ or 0.6 (0.45 or 0.55) for 
$\sin^2{2\theta_{23}}=0.96\,(0.99)$.

For the appearance channel, we use the 
$\nu_{\mu} (\bar{\nu}_\mu) \to \nu_{e} (\bar\nu_e$) 
oscillation probability with first-order matter effect \cite{AKS}
\begin{eqnarray}
P[\nu_{\mu}(\bar{\nu}_{\mu}) &\rightarrow&  
\nu_{\rm e}(\bar{\nu}_e)]  =  
c^2_{23} \sin^2{2\theta_{12}} 
\left(\frac{\Delta m^2_{21} L}{4 E}\right)^2 
\nonumber \\
&+& \sin^2{2\theta_{13}} s^2_{23}
\left[
\sin^2 \left(\frac{\Delta m^2_{31} L}{4 E}\right)
-\frac {1}{2}
s^2_{12}
\left(\frac{\Delta m^2_{21} L}{2 E}\right)
\sin \left(\frac{\Delta m^2_{31} L}{2 E}\right) 
\right.
\nonumber \\
&&\hspace*{22mm} {}\pm
\left.
\left(\frac {4 Ea}{\Delta m^2_{31}}\right)
\sin^2 {\left(\frac{\Delta m^2_{31} L}{4 E}\right)}
\mp 
\frac{aL}{2}
\sin \left(\frac{\Delta m^2_{31} L}{2 E}\right) 
\right]
\nonumber \\
&+& 
2J_{r} \left(\frac{\Delta m^2_{21} L}{2 E} \right)
\left[
\cos{\delta}
\sin \left(\frac{\Delta m^2_{31} L}{2 E}\right) \mp 
2 \sin{\delta}
\sin^2 \left(\frac{\Delta m^2_{31} L}{4 E}\right) 
\right], 
\label{Pmue}
\end{eqnarray}
where the terms of order 
$s_{13} \left( \frac{\Delta m^2_{21}}{\Delta m^2_{31}} \right)^2$ 
and 
$aL s_{13} \left( \frac{\Delta m^2_{21}}{\Delta m^2_{31}} \right)$ 
are neglected. 
In Eq.~(\ref{Pmue}), $a\equiv \sqrt 2 G_F N_e$  \cite{MSW}
where $G_F$ is the Fermi constant, $N_e$ denotes the averaged 
electron number density along the neutrino trajectory in 
the earth, 
$J_r$ $(\equiv c_{12} s_{12} c_{13}^2 s_{13} c_{23} s_{23} )$ 
denotes the reduced Jarlskog factor, and the upper and the 
lower sign $\pm$ refer to the neutrino and 
anti-neutrino channels, respectively. 
We take constant matter density approximation in this paper.
The first term of Eq.~(\ref{Pmue}) is due to the oscillation driven by the 
solar $\Delta m^2_{21}$, which is essentially negligible in the intermediate 
detector but not at the far detector and is of key importance to resolve the 
$\theta_{23}$ octant degeneracy.

We make an approximation of ignoring terms of order 
$(\Delta m^2_{21}/\Delta m^2_{31}) J_r \cos{2\theta_{23}}$ in 
Eq.~(\ref{Pmue}). 
Note that keeping only the leading order in this quantity 
is reasonable because $J_r \lsim 0.04$, 
$|\Delta m^2_{21}/\Delta m^2_{31}| \simeq 1/30$, and 
$\cos{2\theta_{23}}= \pm 0.2$ for $\sin^2{2\theta_{23}} = 0.96$. 
Then, 
the two degenerate solutions obey an approximate relationship 
\begin{eqnarray}
\left( \sin^2 {2\theta_{13}} s^2_{23}  \right)^{\text{1st}} =
\left( \sin^2 {2\theta_{13}} s^2_{23}  \right)^{\text{2nd}}, 
\label{1st_order}
\end{eqnarray}
or, 
$s_{13}^{\text{1st}}s_{23}^{\text{1st}} 
= s_{13}^{\text{2nd}}s_{23}^{\text{2nd}}$
ignoring higher order terms in $s_{13}$. 
We can neglect the leading order correction in $s^2_{13}$ to 
$s^2_{23}$ in these relations because it gives $O(s^4_{13})$ terms.

Analytic treatment of the intrinsic and the sign-$\Delta m^2$ degeneracies 
is given in \cite{MNP2}. 
In an environment where the vacuum oscillation approximation applies the 
solutions corresponding to the intrinsic degeneracy are given by \cite{intrinsic} 
\begin{eqnarray}
\theta_{13}^{(2)} = \theta_{13}^{(1)}, 
\hspace{1cm}
\delta^{(2)} = \pi -  \delta^{(1)},
\label{intrinsicDS}
\end{eqnarray}
where the superscripts (1) and (2) label the solutions due to the 
intrinsic degeneracy. 
Under the same approximation the solutions corresponding to the 
sign-$\Delta m^2$ degeneracy are given by \cite {MNjhep01}
\begin{eqnarray}
\theta_{13}^{\,\text{norm}} = \theta_{13}^{\,\text{inv}}, 
\hspace{1cm}
\delta^{\,\text{norm}} = \pi - \delta^{\,\text{inv}}, 
\hspace{1cm}
(\Delta m^2_{31})^{\,\text{norm}} = - (\Delta m^2_{31})^{\,\text{inv}}, 
\label{signdm2DS}
\end{eqnarray}
where the superscripts ``norm'' and ``inv'' label the solutions with 
the positive and the negative sign of $\Delta m^2_{31}$.  
The degeneracy stems from the approximate symmetry under the 
exchange of these two solutions through which the degeneracy is 
uncovered \cite {MNjhep01}.
The validity of these approximate relationships  in the actual experimental 
setup in the T2K II measurement is explicitly verified in \cite{T2KK}. 
It should be noticed that even if sizable matter effect is present 
the relation (\ref{intrinsicDS}) holds 
at the energy corresponding to the vacuum oscillation maximum, 
or more precisely, the shrunk ellipse limit \cite{KMN02}.

\subsection{Decoupling between degeneracies}

Resolution of the degeneracy can be done when a measurement distinguishes 
between the values of the oscillation probabilities with the two different 
solutions corresponding to a degeneracy. 
Therefore, we define the probability difference 
\begin{eqnarray}
&&\Delta P^{ab}(\nu_{\alpha} \rightarrow \nu_{\beta}) 
\nonumber \\
&\equiv&  
P \left( \nu_{\alpha} \rightarrow \nu_{\beta}; \theta_{23}^{(a)}, 
\theta_{13}^{(a)}, \delta^{(a)}, (\Delta m^2_{31})^{(a)} \right) - 
P \left( \nu_{\alpha} \rightarrow \nu_{\beta}; \theta_{23}^{(b)}, 
\theta_{13}^{(b)}, \delta^{(b)}, (\Delta m^2_{31})^{(b)} \right)
\label{DeltaPdef}
\end{eqnarray}
where the superscripts $a$ and $b$ label the degenerate solutions. 
Suppose that we are discussing the degeneracy A. 
The decoupling between the degeneracies A and B  
holds if $\Delta P^{ab}$ defined in (\ref{DeltaPdef}) for the degeneracy A 
is invariant under the replacement of the mixing parameters 
corresponding to the degeneracy B, and vice versa.

The best example of the decoupling is given by the one between 
the $\theta_{23}$ octant and the sign-$\Delta m^2$ degeneracies. 
By noting that 
$J_r^{\,\text{1st}} - J_r^{\,\text{2nd}} 
= \cos 2\theta_{23}^{\,\text{1st}} J_r^{\,\text{1st}}$ 
in leading order in $\cos{2\theta_{23}}$, 
the difference between probabilities with the first and the second octant 
solutions can be given by 
\begin{eqnarray}
&&\Delta P^{\,\text{1st 2nd}}(\nu_{\mu} \rightarrow \nu_{e}) = 
\cos{2\theta_{23}^{\,\text{1st}}} 
\sin^2{2\theta_{12}} 
\left(\frac{\Delta m^2_{21} L}{4 E}\right)^2 
\nonumber \\
&+&
2J_{r}^{\,\text{1st}} \cos 2\theta_{23}^{\,\text{1st}} 
\left(\frac{\Delta m^2_{21} L}{2 E} \right)
\left[
\cos{\delta}
\sin \left(\frac{\Delta m^2_{31} L}{2 E}\right) \mp 
2 \sin{\delta}
\sin^2 \left(\frac{\Delta m^2_{31} L}{4 E}\right) 
\right]. 
\label{DeltaP_23}
\end{eqnarray}
The remarkable feature of (\ref{DeltaP_23}) is that the leading-order 
matter effect terms drops out completely. 
Therefore, our approximated treatment remains valid until the 
second order matter effect starts to become sizable in the 
appearance oscillation probability. 
More importantly, $\Delta P^{\,\text{1st~2nd}}$, being composed only of 
the vacuum oscillation terms, 
is obviously invariant under the replacement 
$normal \leftrightarrow inverted$ solutions with different signs of 
$\Delta m^2_{31}$ given in Eq.~(\ref{signdm2DS}). 
Therefore, the resolution of the $\theta_{23}$ octant degeneracy 
can be carried out without worrying about the presence of 
the sign-$\Delta m^2_{31}$ degeneracy.

Next, we examine the inverse problem; 
Does the determination of mass hierarchy decouple with 
the resolution of the $\theta_{23}$ degeneracy?
One can show, by using Eq.~(\ref{Pmue}), that 
the similar probability difference between the solutions in 
Eq.~(\ref{signdm2DS}) with the normal and the 
inverted hierarchies is given by 
\begin{eqnarray}
&&\Delta P^{\,\text{norm~inv}} (\nu_{\mu} \rightarrow \nu_{e}) 
\nonumber \\
&=&
\sin^2{2\theta_{13}^{\,\text{norm}}} (s_{23}^{\,\text{norm}})^2
\left[ 
- s^2_{12}
\left(\frac{\Delta m^2_{21} L}{2 E}\right)
\sin \left(\frac{(\Delta m^2_{31})^{\,\text{norm}} L}{2 E}\right) 
\right.
\nonumber \\
\hspace*{0mm} {}&\pm&
\left.
2(aL) 
\left\{
 \left(\frac{4E}{(\Delta m^2_{31})^{\,\text{norm}} L}\right) 
\sin^2 \left(\frac{(\Delta m^2_{31})^{\,\text{norm}} L}{4 E}\right) -
\frac{1}{2} \sin \left(\frac{(\Delta m^2_{31})^{\,\text{norm}} L}{2 E}\right) 
\right\}
\right]
\label{DeltaP_sign}
\end{eqnarray}
where the superscripts ``norm'' and  ``inv'' can be exchanged 
if one want to start from the inverted hierarchy. 
We notice that most of the vacuum oscillation terms, including the solar term, 
drop out because of the invariance under 
$\delta \rightarrow \pi - \delta$ and 
$\Delta m^2_{31} \rightarrow - \Delta m^2_{31}$. 
Now, we observe that $\Delta P^{\,\,\text{norm~inv}}$
is invariant under the transformation 
$\theta_{23}^{\,\text{1st}} \leftrightarrow \theta_{23}^{\,\text{2nd}}$ and 
$\theta_{13}^{\,\text{1st}} \leftrightarrow \theta_{13}^{\,\text{2nd}}$, 
because $\Delta P^{\,\,\text{norm~inv}}$ depends upon 
$\theta_{13}$ and $\theta_{23}$ only through the combination 
$\sin^2{2\theta_{13}} s_{23}^{2}$ within our approximation. 
Therefore, the sign-$\Delta m^2_{31}$ and the $\theta_{23}$ 
degeneracies decouple with each other.

In our previous paper, we have shown that the sign-$\Delta m^2_{31}$ 
degeneracy can be lifted by the Kamioka-Korea two detector setting. 
The above argument for decoupling guarantees that our treatment 
is valid irrespective of the solutions assumed for $\theta_{23}$ 
degeneracy.\footnote{
We remark that in most part of \cite{T2KK}, we have assumed that 
$\theta_{23}=\pi/4$ so that this problem itself does not exist. 
The above discussion implies that even in the case of 
non-maximal value of $\theta_{23}$ the similar analysis 
as in \cite{T2KK} must go through without knowing in which octant 
$\theta_{23}$ lives. 
The resultant sensitivity for resolving the mass hierarchy will change 
as $\theta_{23}$ goes away from $\pi/4$ but only slightly 
as will be shown in Sec.~\ref{sensitivity}. 
}

\subsection{Including the intrinsic degeneracy}
\label{intrinsic}

Now we turn to the intrinsic degeneracy for which the situation 
is somewhat different. First of all, in our setting, 
the intrinsic degeneracy is already 
resolved by spectrum informations at the intermediate detector 
if $\theta_{13}$ is relatively large, $\sin^2 2\theta_{13} \gsim 0.02$ 
as illustrated in Fig.~\ref{intrinsicKamKorea} before the information 
from the far detector is utilized. 
It means that there is no intrinsic degeneracy from the beginning 
in the analysis with spectrum informations.  
Because of this feature, the intrinsic degeneracy decouple from 
the beginning from the task of resolving the eight-fold degeneracy in 
our setting for the relatively large values of $\theta_{13}$.
Because of a further enhanced sensitivity, the intrinsic degeneracy 
may be resolved to a smaller values of $\theta_{13}$ in the Kamioka-Korea setting.

In fact, powerfulness of the spectral information for resolving 
the intrinsic degeneracy can be understood easily by noting that 
$\Delta P^{ab}$ defined in (\ref{DeltaPdef}) is given by using the 
intrinsic degeneracy solution (\ref{intrinsicDS}) as 
\begin{eqnarray}
&&
\Delta P^{12}(\nu_{\mu} \rightarrow \nu_{e}) =
4 J_{r} \left(\frac{\Delta m^2_{21} L}{2 E} \right) 
\cos{\delta^{(1)}}
\sin \left(\frac{\Delta m^2_{31} L}{2 E}\right),
\label{DeltaP_intrinsic}
\end{eqnarray}
with notable feature that the matter effect cancels out. 
Notice that even if the matter effect cannot be negligible the solution 
(\ref{intrinsicDS}) holds for measurement at energies around the 
vacuum oscillation maximum. 
The right-hand side of Eq.~(\ref{DeltaP_intrinsic}) is proportional to $E^{-2}$ 
at energies near the vacuum oscillation maximum, 
$\frac{\Delta m^2_{21} L}{2 E} = \pi$, and the steep energy 
dependence can be used to lift the degeneracy. 
Hence, the spectrum analysis is a powerful tool for resolving the 
intrinsic degeneracy.

\subsection{The case that the intrinsic degeneracy is not solved}

Even if $\theta_{13}$ is too small, or if the energy resolution is too 
poor for the spectrum analysis to resolve the intrinsic degeneracy, 
we can show that  the intrinsic degeneracy approximately 
decouples from the other degeneracies. 
$\Delta P^{12}$ in Eq.~(\ref{DeltaP_intrinsic}) is not exactly but approximately 
invariant under the transformation first-octant $\leftrightarrow$ 
second-octant solutions. 
The difference between $\Delta P^{12} (\text{1st})$ and 
$\Delta P^{12} (\text{2nd})$
is of order $\cos 2\theta_{23} J_r \Delta m^2_{21} / \Delta m^2_{31} 
\simeq 3 \times 10^{-4}$ for $s^2_{23}=0.4$ and $\sin^2 2\theta_{13}= 0.1$
apart from the further suppression by 
$\sin \left(\frac{\Delta m^2_{31} L}{2 E}\right)$ at around the 
oscillation maximum. 
Being the vacuum oscillation term $\Delta P^{12}$ is obviously invariant 
under the replacement $normal \leftrightarrow inverted$ solutions 
with different signs of $\Delta m^2_{31}$. 
Therefore, resolution of the intrinsic degeneracy can be done, 
to a good approximation, independent of the presence of the 
sign-$\Delta m^2_{31}$ and the $\theta_{23}$ octant degeneracies.

The remaining problem we need to address is the inverse problem, 
whether the resolution of the sign-$\Delta m^2_{31}$ and the $\theta_{23}$ 
octant degeneracies can be carried out without knowing solutions 
of the intrinsic degeneracy. 
The sign-$\Delta m^2_{31}$ degeneracy decouples from the intrinsic one 
because $\Delta P^{\,\text{norm~inv}}$ in (\ref{DeltaP_sign}) is invariant under 
the exchange of two intrinsic degeneracy solutions. 
The $\theta_{23}$ octant degeneracy also approximately decouples 
from the intrinsic one. 
$\Delta P^{\,\text{1st~2nd}}$ in (\ref{DeltaP_23}) changes under the 
interchange of two intrinsic degeneracy solutions only by the 
same amount as the difference between $\Delta P^{12}$ of 
the first and the octant $\theta_{23}$ solutions,  
$\cos 2\theta_{23} J_r \Delta m^2_{21} / \Delta m^2_{31} 
\simeq 3 \times 10^{-4}$ (for $s^2_{23}=0.4$ and $\sin^2 2\theta_{13}= 0.1$).

Here is a clarifying comment on what the decoupling really means; 
Because of the cross-shaped structure of the degenerate solutions 
of the intrinsic and the sign-$\Delta m^2_{31}$ degeneracies 
(as was shown in Sec.~\ref{how}) 
the decoupling of the former from the latter does {\em not} imply that 
the correct value of $\delta$ can be extracted from the measurement 
without knowing the correct sign of $\Delta m^2_{31}$.\footnote{
Notice, however, that it does {\em not} obscure the CP violation, 
because the ambiguity is only two-fold; $\delta \leftrightarrow \pi-\delta$. 
}
It means that the elimination of one of the ``intrinsic'' degenerate pair 
solutions related by $\delta \leftrightarrow \pi-\delta$ 
for a given sign of $\Delta m^2_{31}$ can be done without knowing 
the mass hierarchy, the true sign of $\Delta m^2_{31}$. 
Therefore, the situation that the intrinsic degeneracy is always 
resolved by the spectrum analysis in region of not too small 
$\theta_{13}$, as is the case in our setting, is particularly transparent 
one from this viewpoint.

To sum up, we have shown that to leading order in the matter effect  
the intrinsic, the sign-$\Delta m^2_{31}$, and the $\theta_{23}$ 
octant degeneracies decouples with each other. 
They do so exactly except for between the intrinsic and the $\theta_{23}$ 
octant degeneracies for which the decoupling is approximate but 
sufficiently good to allow one-by-one resolution of all the three types 
of degeneracies. 
The decoupling implies that in analysis for lifting the eight-fold 
degeneracy the structure of the $\chi^2$ minimum is very simple in 
multi-dimensional parameter space, and it may be of use in 
discussions of how to solve the degeneracy  
in much wider context than that discussed in this paper.

\section{How identical two detector setting solves $\theta_{23}$ octant degeneracy?}
\label{23degeneracy}

Now, we turn to the problem of how the identical two detector 
setting can resolve the $\theta_{23}$ degeneracy, 
the unique missing link in a program of resolving eight-fold 
parameter degeneracy in the Kamioka-Korea two detector setting. 
The solar $\Delta m^2$ oscillation term, 
the first term in Eq.~(\ref{Pmue}) with the coefficient of $c^2_{23}$, 
may be of key importance to do the job. 
While it was argued on very general ground \cite{resolve23} that
the $\theta_{23}$ degeneracy is hard to resolve only by accelerator 
experiments with baseline of $\lsim 1000$ km or so,  
the argument can be circumvented if the solar term can be isolated. 
We emphasize that the accuracy of the determination of $\theta_{23}$ is 
severely limited by the octant degeneracy, as discussed in detail 
in \cite{MSS04}.

Therefore, the question we must address first is the relative importance 
of the solar term to the remaining terms in $\Delta P^{\,\text{1st~2nd}}$ 
in Eq.~(\ref{DeltaP_23}). 
We note that the ratio of the solar term to the $\delta$-dependent 
solar-atmospheric interference term in $\Delta P^{\,\text{1st~2nd}}$ is 
given by 
$\sin^2{2\theta_{12}} (\Delta m^2_{21} L / 4 E) / 4 J_{r}$, 
assuming the square parenthesis in (\ref{DeltaP_23}) is of order unity. 
The ratio is roughly given by 
$\simeq 3 (1 / 30) (\pi / 2) 0.86 (1/ 4 J_{r} ) \simeq 0.9~(0.16 / s_{13}) $ 
with beam energy having the first oscillation maximum in Kamioka. 
%
Therefore, the solar term is indeed comparable or lager for
smaller $\theta_{13}$ in size with the interference terms 
in $\Delta P^{\,\text{1st~2nd}}$ at the far detector. 
Obviously, the solar term is independent of $\theta_{13}$,
which suggests that the sensitivity to resolve the $\theta_{23}$
degeneracy is almost independent of $\theta_{13}$, 
as will be demonstrated in Sec.~\ref{sensitivity}. 
We note that while the solar term is the key to resolve 
the $\theta_{23}$ degeneracy, the interference terms 
also contributes to lift the degeneracy. In particular, 
as shown in (\ref{DeltaP_23}), 
the $\sin\delta$ term has opposite sign when the polarity of the 
beam is switched from the neutrino to the anti-neutrino runs.

\begin{figure}[htbp]
\begin{center}
\includegraphics[width=0.8\textwidth]{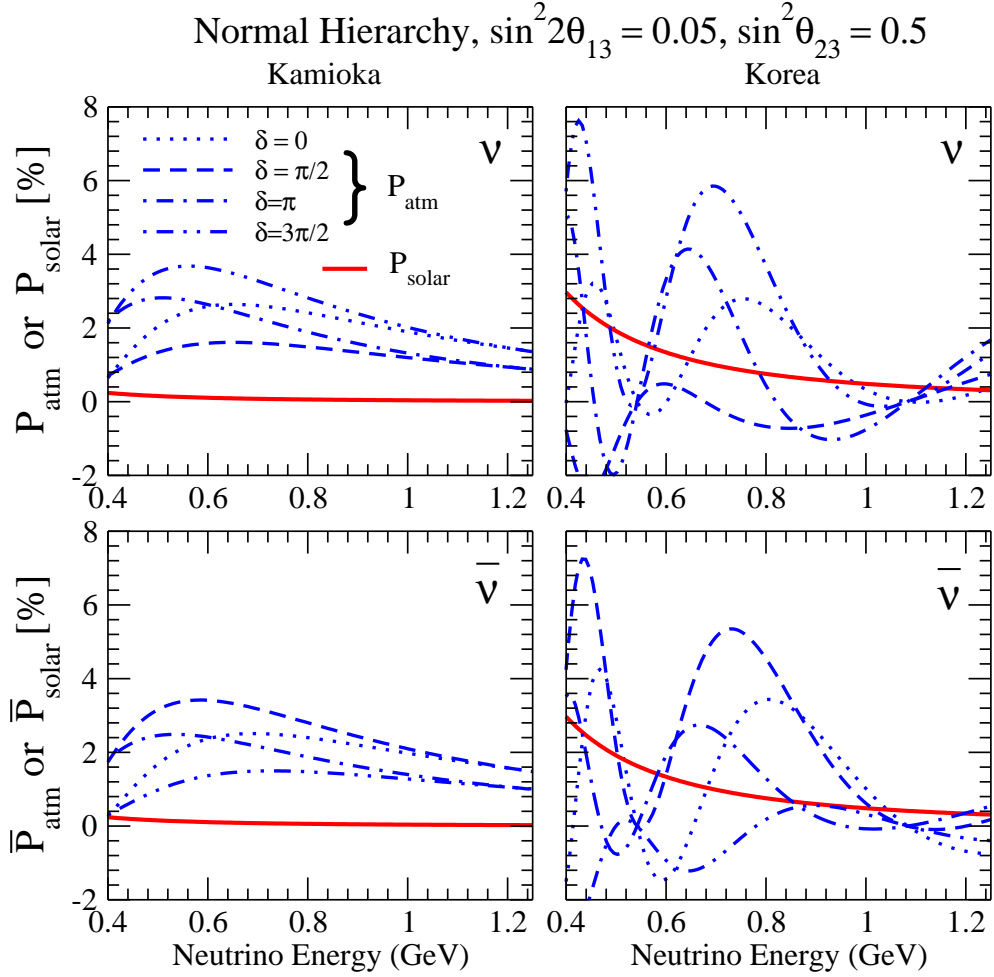}
\end{center}
\caption{
The energy dependence of the solar term (red solid line) is 
contrasted with the ones of atmospheric plus interference 
terms in the $\nu_{e}$ 
appearance oscillation probabilities with various values of CP phase 
$\delta$; 
$\delta=0$ (dotted line), 
$\delta=\pi/2$ (dashed line), 
$\delta=\pi$ (dash-dotted line), and 
$\delta=3\pi/2$ (double-dash-dotted line). 
For this plot, we used analytic expression in Eq.~(\ref{Pmue}); 
$P_{\text{solar}}$ is defined to be the first term in 
Eq.~(\ref{Pmue}) whereas $P_{\text{atm}}$ is defined to be 
the rest in Eq.~(\ref{Pmue}).  
$\bar{P}_{\text{solar}}$ and $\bar{P}_{\text{atm}}$ refer to the 
corresponding terms for anti-neutrinos. 
}
\label{solar-vs-atm}
\end{figure}

The next question we must address is how the solar term can be 
separated from the other terms to have enhanced sensitivity to 
the $\theta_{23}$ degeneracy. 
To understand the behavior of the solar term and its difference from 
that of the atmospheric terms in the oscillation probability, 
we plot in Fig.~\ref{solar-vs-atm} a comparison between them 
in Kamioka (left panels) and in Korea (right panels) for various 
values of $\delta$.
As one can observe in the right panels, the energy dependence 
of the solar oscillation term,  a monotonically decreasing 
(approximately $1/E^2$) behavior with increasing energy, 
is quite different  from the oscillating behavior of the atmospheric ones. 
It is also notable that the ratio of the solar term to the atmospheric-solar 
interference term is quite different between the intermediate and 
the far detectors.
Due to the differing relative importance of the solar term in the two 
detectors and the clear difference in the energy dependences 
between the solar and the atmospheric terms, 
the spectrum analysis, the powerful method for resolving the 
intrinsic degeneracy, must be able to isolate the solar term 
from the remaining ones. 
This will be demonstrated in the quantitative analysis in the next section.

We note that several alternative methods are proposed to resolve 
the $\theta_{23}$ degeneracy. 
They include: 
the atmospheric neutrino method \cite{concha-smi_23,atm23,choubey2}, and 
the reactor accelerator combined method \cite{MSYIS,resolve23}, 
the atmospheric accelerator combined method \cite{atm-lbl}. 
The atmospheric neutrino method discussed in 
\cite{concha-smi_23,atm23} is closest to ours in physics 
principle of utilizing the solar mass scale oscillation effect. 
Possible advantage of the present method may be in a clean detection 
of the solar term by the intermediate-versus-far two detector comparison.

\section{Sensitivity for resolving  $\theta_{23}$ octant degeneracy}
\label{sensitivity}

In this section, we describe details of our analysis for resolving 
the $\theta_{23}$ octant degeneracy. 
They include treatment of experimental errors, treatment of 
background, and the statistical procedure which is used to 
investigate the sensitivity of the experiment. 
Then, the results of our analysis are presented.

\subsection{Assumptions and the definition of $\chi^2$}

In order to understand the sensitivity of the experiment
with the two detector system at 
295~km (Kamioka) and 1050~km (Korea), we carry out 
a detailed $\chi^2$ analysis. 
To address the $\theta_{23}$ octant degeneracy, 
it is of course necessary to include $\nu_{\mu}$ and $\bar{\nu}_{\mu}$ 
disappearance channels in addition to the appearance ones in our treatment. 
In short, the definition of the statistical procedure is similar to the 
one used in Ref.~\cite{T2KK} with necessary extension for including 
muon events. 
The assumption on the experimental setting is also identical
to that of the best performance setting identified in Ref.~\cite{T2KK}.
Namely, 0.27~Mton fiducial masses for the intermediate 
site (Kamioka, 295~km) and the far site (Korea, 1050~km).
The neutrino beam is assumed to be 2.5 degree off-axis one produced
by the upgraded J-PARC 4~MW proton beam. 
It is assumed that the experiment will continue for 8 years
with 4 years of neutrino and 4 years of anti-neutrino runs. 

We use various numbers and distributions available from
references related to T2K  \cite{JPARC-detail}, in which many of 
the numbers are updated after the original proposal \cite{T2K}.
Here, we summarize the main assumptions and the methods 
used in the $\chi^2$ analysis.
We use the reconstructed neutrino energy for single-Cherenkov-ring 
electron and muon events.
The resolution in the reconstructed neutrino energy is 80~MeV 
for quasi-elastic events. 
We assume that 
$|\Delta m_{31}^2|$ should be known precisely by the time when 
the experiment we consider in this report will be carried out. 
We take 
$\Delta m_{31}^2 = \pm 2.5 \times 10^{-3} \text{eV}^2$. 
Hence, we assume that the 
energy spectrum of the beam is the one expected by the
2.5~degree off-axis-beam in T2K. The shape of the
energy spectrum for the anti-neutrino beam is assumed to be
identical to that of the neutrino beam. The event rate
for the anti-neutrino beam 
in the absence of neutrino oscillations
is smaller by a factor of 3.4 due mostly to the lower neutrino 
interaction cross sections and partly to the slightly lower 
flux. The signal to noise ratio is
worse for the anti-neutrino beam than that for the neutrino beam
by a factor of about 2. 

28 background electron events are expected 
for the reconstructed neutrino energies between 350 and 850~MeV for
$(0.75 \times 0.0225 \times 5)\, \text{MW} \cdot \text{Mton} \cdot \text{yr}$ 
measurement 
with the neutrino beam. The energy dependence 
of the background rate and the rate itself are taken 
from \cite{JPARC-detail}.
The background rate is expected to be higher in the lower 
neutrino energies. The expected number of electron events
is assumed to be 122 for $\sin^2 2\theta_{13} =$0.1 with the 
same detector exposure
and beam, assuming the normal mass hierarchy and $\delta = 0$.

We assume that the experiment is equipped with a near detector 
which measures the rate and the energy dependence of the background
for electron events,
un-oscillated muon spectrum, 
and the signal detection efficiency.
These measurements are assumed to be carried out
within the uncertainty of 5\%. 
We already demonstrated that the dependence on the 
assumed value of the experimental systematic errors is 
rather weak \cite{T2KK}.
We stress that in the present setting 
the detectors located in Kamioka and in Korea are not only 
identical but also receive neutrino beams with essentially
the same energy distribution (due to the same off-axis angle of 2.5 degree) 
in the absence of oscillations. 
However, it was realized recently that, due to a non-circular shape 
of the decay pipe of the J-PARC neutrino beam line, 
the flux energy spectra viewed at detectors in Kamioka and in 
Korea are expected to be slightly different even at the same off-axis angle, 
especially in the high-energy tail of the spectrum 
\cite{Rubbia-Meregaglia-Seoul2006}. 
The possible difference between fluxes in the intermediate and the 
far detectors is newly taken into account as a systematic error in 
the present analysis.

We compute neutrino oscillation probabilities 
by numerically integrating neutrino evolution equation 
under the constant density approximation.  
The average density is assumed to be 2.3 and 2.8 g/cm$^3$ for the 
matter along the beam line between the production target and 
Kamioka and between the target and Korea, respectively \cite{T2KK}.
We assume that the number of electron with respect to that of nucleons 
to be 0.5 to convert the matter density to the electron number density. 
In our $\chi^2$ analysis, 
we fix the absolute value of $|\Delta m_{31}^2|$ to be $2.5\times 10^{-3}$ eV$^2$, 
and fix solar parameters as $\Delta m_{21}^2 = 8\times 10^{-5}$ eV$^2$ and 
$\sin^2{\theta_{12}}$=0.31.

\begin{figure}[htbp]
\vglue 0.3cm
\begin{center}
\includegraphics[width=0.8\textwidth]{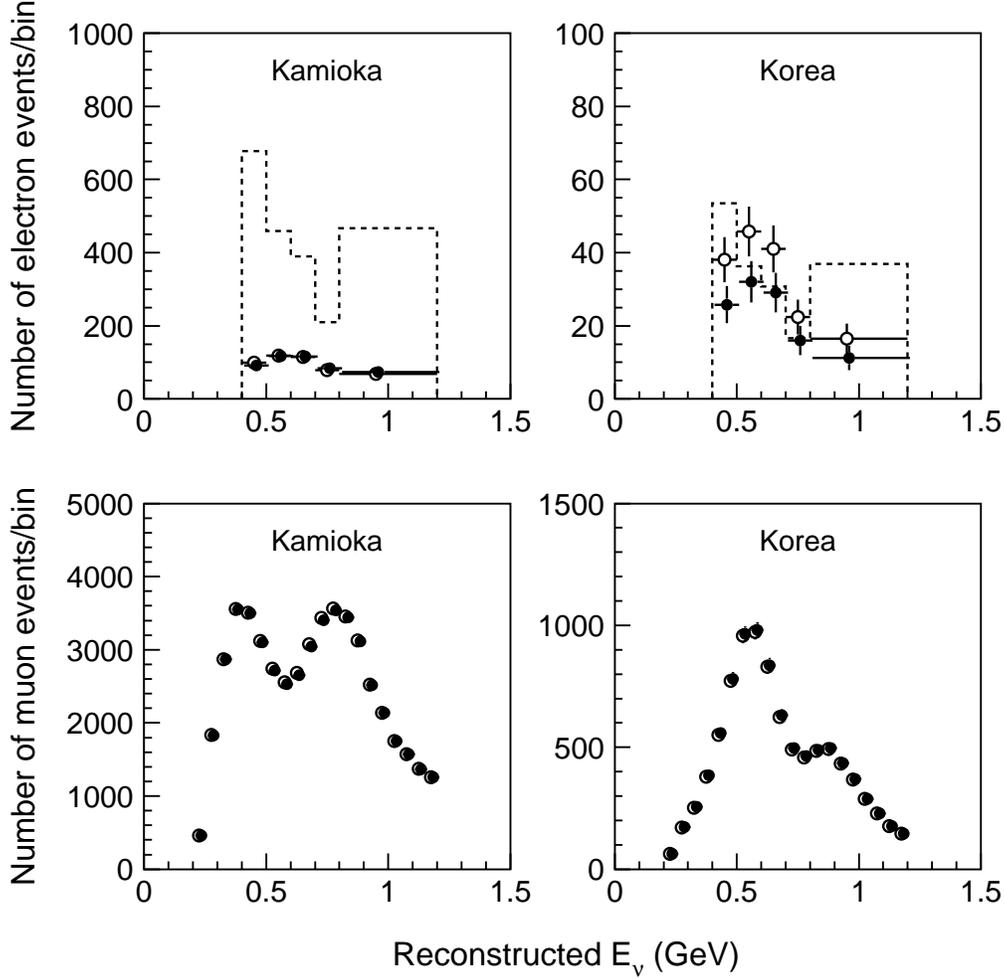}
\end{center}
\vglue -0.3cm
\caption{Examples of electron and muon events to be observed in 
Kamioka and Korea for 4 years of neutrino plus 
4 years of anti-neutrino running 
are presented as a function of reconstructed neutrino energy. 
The fiducial masses are taken to be 0.27~Mton for both the detectors in 
Kamioka and Korea.
The dashed histograms for electron events show the background events. 
The open circles show the expected
energy spectrum of signal events with 
$\sin^2 \theta_{23} =$0.40 and $\sin^2 2 \theta_{13} =$0.01.
The solid circles show the expected
energy spectrum of signal events with 
$\sin^2 \theta_{23} =$0.60 and $\sin^2 2 \theta_{13} =$0.0067.
In both cases, $\delta = 3\pi/4$ and normal mass
hierarchy are assumed in simulating the events. 
}
\label{fig:energy-spectrum-examples}
\end{figure}

Fig.~\ref{fig:energy-spectrum-examples} shows an
example of the energy spectrum of electron and muon events
to be observed in Kamioka and Korea for 4 years of
neutrino beam plus 4 years of anti-neutrino beam.
The two sets of parameters give very similar spectrum 
for both the electron and muon events at the Kamioka detector 
and the muon events at the Korean detector. 
However, due to the long baseline distance, the
solar term plays some role in the 
$^( \overline{\nu}^)_\mu \rightarrow\, ^(\overline{\nu}^)_e$ 
oscillation 
probability at the Korean detector. Therefore, 
the two sets of parameters give slightly different
oscillation probabilities in Korea. 
Since the solar term is proportional to $c^2_{23}$ 
we use this feature to obtain information on
$\sin^2 \theta_{23}$.

The statistical significance of the measurement considered in this
paper was estimated by using the following 
definition of $\chi^2$:
\begin{equation}
\chi^2 =  \sum_{k=1}^{4} \left( 
\sum_{i=1}^{5}
\frac{\left(N(e)_{i}^{\rm obs} - N(e)_{i}^{\rm exp}\right)^2}
{ \sigma^2_{i} } +
\sum_{i=1}^{20}
\frac{\left(N(\mu)_{i}^{\rm obs} - N(\mu)_{i}^{\rm exp}\right)^2}
{ \sigma^2_{i} }
\right)
+ \sum_{j=1}^{7} \left(\frac{\epsilon_j}
{\tilde{\sigma}_{j}}\right)^2, 
\label{equation:chi2def}
\end{equation}
where
\begin{eqnarray}
 N(e)_{i}^{\rm exp} = N_{i}^{\rm BG} \cdot 
   (1+\sum_{j=1,2,7} f(e)_{j}^{i}\cdot\epsilon_{j}) 
    + N_i^{\rm signal} \cdot 
   (1+\sum_{j=3,7} f(e)_{j}^{i}\cdot\epsilon_{j}) ~,
\label{equation:e-number}
\\
 N(\mu)_{i}^{\rm exp} = N_{i}^{\rm non-QE} \cdot 
   (1+\sum_{j=4,6,7} f(\mu)_{j}^{i}\cdot\epsilon_{j}) 
    + N_i^{\rm QE} \cdot 
   (1+\sum_{j=4,5,7}   f(\mu)_{j}^{i}\cdot\epsilon_{j}) ~.
\label{equation:mu-number}
\end{eqnarray} 

\noindent
%
The first and second terms in Eq.~(\ref{equation:chi2def}) are
 for the number of observed 
single-ring electron and muon events, respectively.
$N(e~{\rm or}~\mu)^{\rm obs}_i$ is the number of events to be 
observed for the given oscillation parameter set,
 and $N(e~{\rm or}~\mu )^{\rm exp}_i$ is the expected number of
events for the assumed oscillation parameters 
in the $\chi^2$ analysis. 
$k=1,2,3$ and $4$ correspond to the four combinations 
of the detectors in Kamioka and in Korea with the 
neutrino and anti-neutrino beams,
respectively.
The index $i$ represents the reconstructed neutrino energy bin 
for both electrons and muons. 
For electron events, both $N(e)^{\rm obs}_i$ and  $N(e)^{\rm exp}_i$ 
include background events.
The energy ranges of the five energy bins for electron events 
are respectively
400-500~MeV, 
500-600~MeV, 600-700~MeV, 700-800~MeV and 
800-1200~MeV. 
The energy range for the muon events covers 
from 200 to 1200~MeV. Each energy bin has 50~MeV width.
$\sigma_i$ denotes the
statistical uncertainties in the expected data. 
The third term  
in the $\chi^2$ definition collects the
contributions from variables which parameterize the systematic
uncertainties in the expected number of signal and background events.

$N^{\rm BG}_i$ is the number of background events
for the $i^{\rm th}$ bin for electrons.
$N^{\rm signal}_i$ is the number of electron appearance events
that are observed, and depends on 
neutrino oscillation parameters.
The uncertainties in $N^{\rm BG}_i$ and $N^{\rm signal}_i$
are represented by 4 parameters $\epsilon_j$ ($j=1$ to 3 and 7).
Similarly, $N^{\rm non-QE}_i$ are the number of non-quasi-elastic events
for the $i^{\rm th}$ bin for muons. 
$N^{\rm QE}_i$ are the number of quasi-elastic muon events.
We treat the non-quasi-elastic and quasi-elastic muon events separately,
since the neutrino energy cannot be properly reconstructed for
 non-quasi-elastic events.
Both  $N^{\rm non-QE}_i$ and $N^{\rm QE}_i$ depend on 
neutrino oscillation parameters.
The uncertainties in  $N^{\rm non-QE}_i$ and $N^{\rm QE}_i$  
are represented by 4 parameters $\epsilon_j$ ($j=4$ to 7).

During the fit, the values of $N(e~{\rm or}~\mu)^{\rm exp}_i$ are
recalculated 
for each choice of the oscillation parameters which are varied freely 
to minimize $\chi^2$, and so are the systematic error parameters 
$\epsilon_j$. 
The parameter $f(e~{\rm or}~\mu)^i_j$ represents 
the fractional change in the predicted
event rate in the $i^{\rm th}$ bin due to a variation of the parameter
$\epsilon_j$. 
The overall background normalization
for electron events 
is assumed to be uncertain by $\pm$5\% ($\tilde{\sigma}_{1}$=0.05). 
      It is also assumed that the background events for electron
      events have an energy dependent uncertainty with the functional
      form of $f(e)_2^i=((E_\nu(rec)-800~\text{MeV}) / 400~\text{MeV})$. 
      5\% is assumed to be the
      uncertainty in $\epsilon_2$ ($\tilde{\sigma}_{2}$=0.05).
      The functional form of
      $f(\mu)_4^i = (E_\nu(rec)-800~\text{MeV}) / 800~\text{MeV}$
      is used to define the uncertainty in the spectrum shape for
      muon events ($\tilde{\sigma}_{4}$=0.05).
The uncertainties in the signal detection efficiency 
are assumed to be 5\% for both electron and muon events
($\tilde{\sigma}_{3}=\tilde{\sigma}_{5}$=0.05).
The uncertainty in the separation of quasi-elastic and 
non-quasi-elastic interactions in the muon events is
assumed to be 20\% ($\tilde{\sigma}_{6}$=0.20).
These systematic errors are assumed to be not correlated between
the electron and muon events.
In addition, for the number of events in Korea, the possible 
flux difference between Kamioka and Korea is taken into account
in $f(e~{\rm or}~\mu)_7^i$. The predicted flux 
difference \cite{Rubbia-Meregaglia-Seoul2006} 
is simply assumed to be the 1~$\sigma$ uncertainty in the flux 
difference ($\tilde{\sigma}_{7}$).

\subsection{Sensitivity with two-detector complex}

\begin{figure}[htbp]
\vglue 0.3cm
\begin{center}
\includegraphics[width=0.9\textwidth]{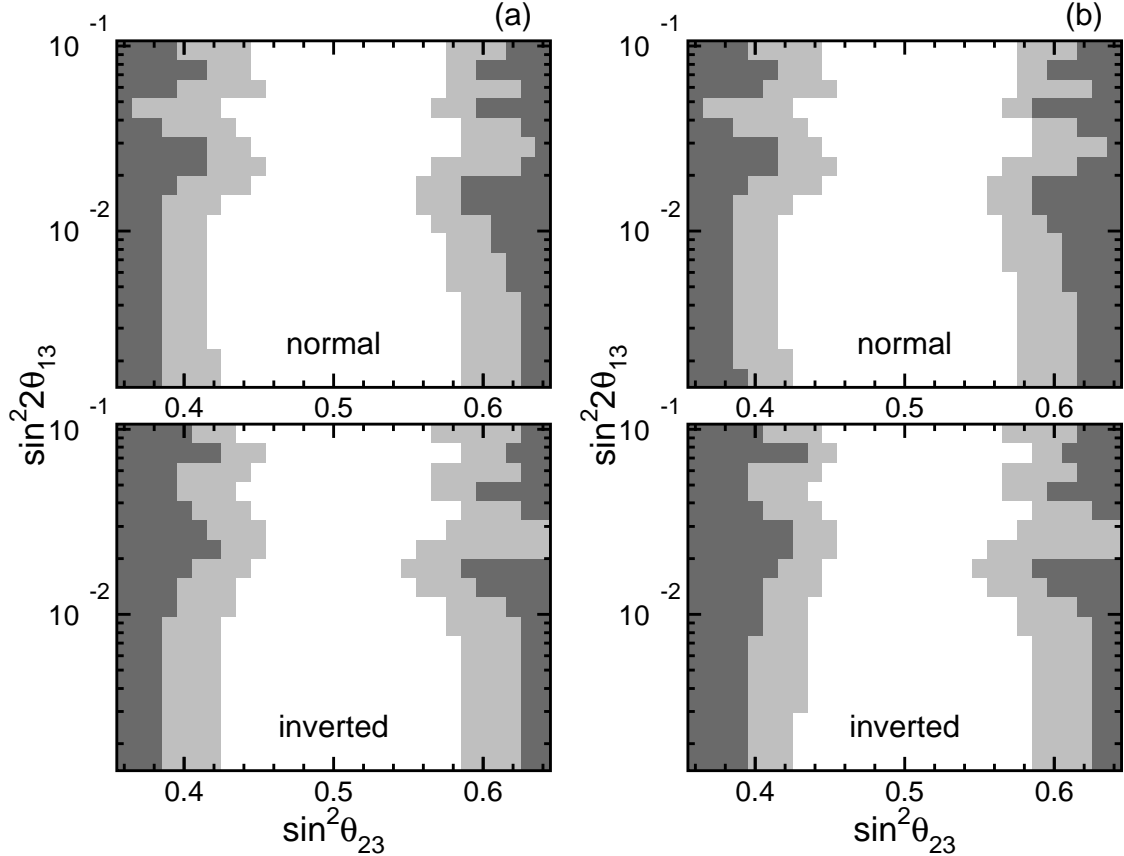}
\end{center}
\vglue -0.3cm
\caption{ 2 (light gray area) and 3 (dark gray area) 
standard deviation sensitivities to the $\theta_{23}$
octant degeneracy for 0.27~Mton detectors both in Kamioka
and Korea.
4 years running with neutrino beam and another 4 years with 
anti-neutrino beam are assumed.
In (a), the sensitivity is defined so that the experiment 
is able to identify the octant of $\theta_{23}$ for any 
values of the CP phase $\delta$. In (b), it is defined so that the 
experiment is able to identify the octant of $\theta_{23}$ 
for half of the CP $\delta$ phase space.
}
\label{sensitivity-theta23-octant}
\end{figure}

Now we present the results of the sensitivity analysis for 
the $\theta_{23}$ octant degeneracy. 
The results for the mass hierarchy as well as CP violation sensitivities 
will be discussed in the next section.
Fig.~\ref{sensitivity-theta23-octant} shows
the sensitivity to the $\theta_{23}$ octant determination as a function of 
$\sin^2 2 \theta_{13}$ and $\sin^2 \theta_{23}$. 
The areas shaded with light (dark) gray of this figure indicate 
the regions of parameters where the octant of $\theta_{23}$ 
can be determined at 2 (3) standard deviation confidence level, 
which is determined by the condition 
$\chi^2_{\text{min}} \text{(wrong~octant)}
-\chi^2_{\text{min}}\text{(true~octant)}>$ 4 (9). 
The upper (lower) panels correspond to the case where
the true hierarchy is normal (inverted). 
Note, however, that the fit was performed without assuming 
the mass hierarchy. 
Since the sensitivity mildly depends on the CP phase $\delta$, 
we define the sensitivity to resolving the octant degeneracy 
in two ways: 
the left (right) panels correspond to the case where 
the sensitivity is defined such that the octant is determined 
for any value of delta (half of the $\delta$ space). 
From this figure, we conclude that the experiment 
we consider here is able to solve the octant ambiguity,
if $\sin^2 \theta_{23} < 0.38\,(0.42) $ or $>0.62\,(0.58)$ at 3
(2) standard deviation confidence level. This conclusion depends
weakly on the value of $\sin^2 2 \theta_{13}$, as well 
as the value of the CP phase $\delta$ and the mass hierarchy.

The sensitivity of lifting the octant degeneracy by this setting 
is quite high even for rather small values of $\theta_{13}$ to 
$\sin^2 2 \theta_{13} \sim 10^{-3}$ where the mass hierarchy is 
not determined, a possible consequence of the decoupling. 
See Figs.~\ref{sensitivity-mass-hierarchy} in the next section. 
The sensitivity depends very weakly on $\theta_{13}$ 
in relatively small values of 
$\sin^2 2\theta_{13}$ where the dominant atmospheric terms 
are small.
The feature of almost independence of the sensitivity to $\theta_{13}$ 
should be contrasted with that of the accelerator-reactor combined 
method in which a strong dependence on $\theta_{13}$ 
is expected \cite{resolve23}. 
Very roughly speaking the sensitivity by the present method is 
better than the latter method in a region 
$\sin^2 2 \theta_{13} \lsim 0.05-0.06$ 
according to the result given in Fig.~8 of \cite{resolve23}. 
The sensitivity of our method is also at least comparable to that could 
be achieved by the high statistics observation of atmospheric neutrinos 
\cite{concha-smi_23,atm23,choubey2}.

\section{Reexamination of sensitivities to neutrino mass hierarchy and CP violation} 
\label{revisit}

In this section we reexamine the problem of sensitivities 
to the neutrino mass hierarchy and CP violation 
achievable by the Kamioka-Korea identical two detector complex.  
We want to verify that the sensitivities do not depend on 
which octant $\theta_{23}$ lives, as indicated by our  
discussion of the decoupling given in Sec.~\ref{decoupling}. 
It is also interesting to examine how the 
sensitivities depend upon $\sin ^2 2\theta_{23}$. 
Furthermore, the inclusion of the new systematic error which accounts 
for difference in the spectral shapes of the neutrino beam between 
the intermediate and the far detectors makes the reexamination 
worth to do.

\begin{figure}[htbp]
\vglue 0.3cm
\begin{center}
\includegraphics[width=0.73\textwidth]{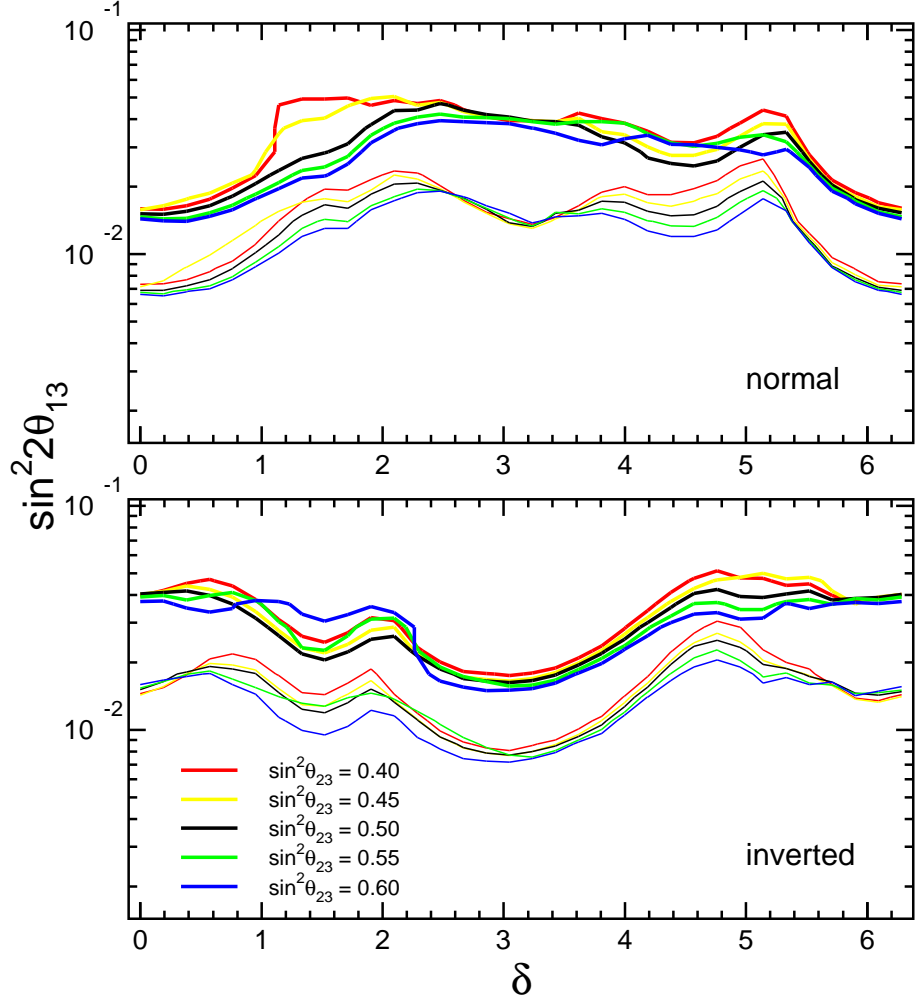}
\end{center}
\vglue -0.3cm
\caption{ 2(thin lines) and 3(thick lines) 
standard deviation sensitivities to the mass hierarchy 
determination for
several values of $\sin ^2 2 \theta_{23}$ (red, yellow, 
black, green and blue lines show the results for 
$\sin^2 \theta_{23} =$ 0.40, 0.45, 0.50, 0.55 and 0.60, 
respectively). The sensitivity is defined in the plane
of $\sin ^2 2 \theta_{13}$ versus CP phase $\delta$.
The top and bottom panels show the cases for positive and 
negative mass hierarchies, respectively. The experimental
setting is identical to that in 
Fig.\ref{sensitivity-theta23-octant}. 
}
\label{sensitivity-mass-hierarchy}
\end{figure}

\begin{figure}[htbp]
\vglue 0.3cm
\begin{center}
\includegraphics[width=0.73\textwidth]{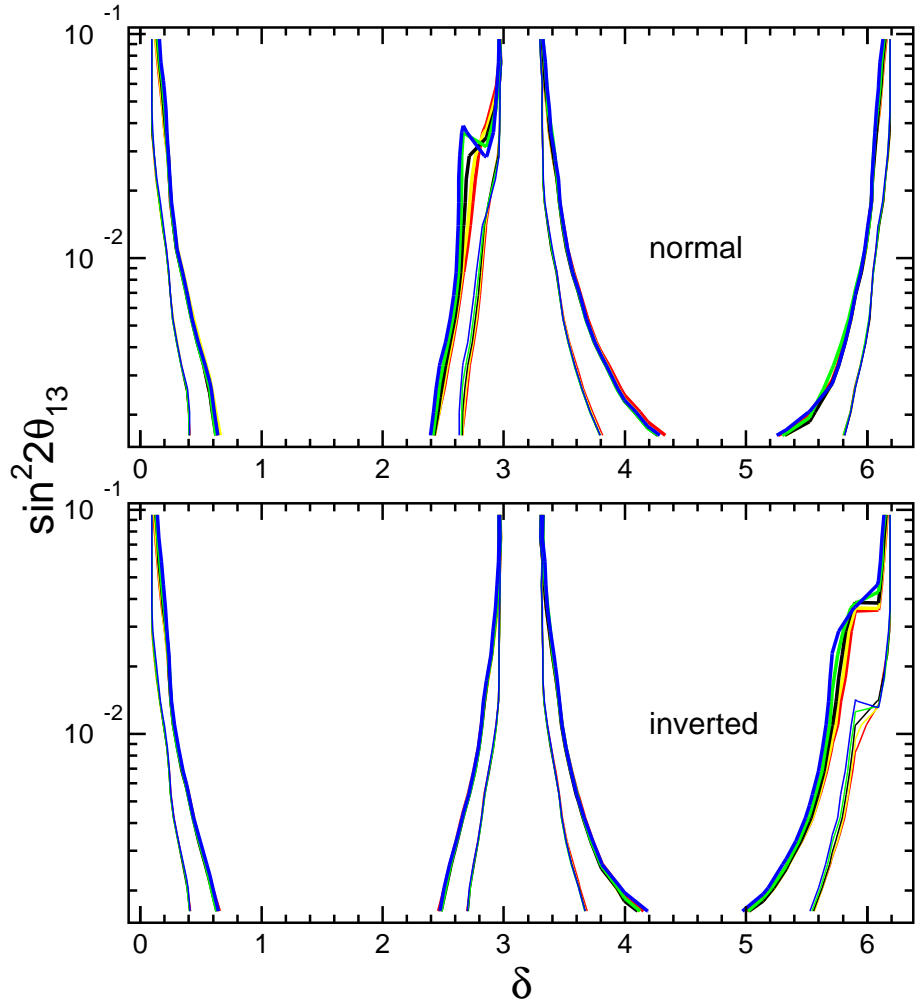}
\end{center}
\vglue -0.3cm
\caption{ Sensitivities to the CP violation, 
$\sin \delta \ne0$. 
The meaning of the lines 
and colors are identical to that in 
Fig.~\ref{sensitivity-mass-hierarchy}.
}
\label{sensitivity-CP}
\end{figure}

In Figs.~\ref{sensitivity-mass-hierarchy} and \ref{sensitivity-CP} 
the regions sensitive to the mass hierarchy and CP violation, 
respectively, are presented. 
In both figures, the thin-lines and the thick-lines 
indicate the sensitivity region at 2 and 3 standard deviations, 
respectively.
As in the previous work~\cite{T2KK}, 2 (3) standard deviation 
sensitivity regions are defined by the conditions, 
$\chi^2_{\text{min}} \text{(wrong~hierarchy)}-\chi^2_{\text{min}}\text{(true~hierarchy)}>$
4 (9) and 
$\chi^2_{\text{min}} (\delta = 0\ \text{or}\ \pi)
-\chi^2_{\text{min}}(\text{true value of}\, \delta)>$
4 (9) for the mass hierarchy and CP violation, respectively.

The sensitivities to the mass hierarchy and CP violation 
at $\sin ^2 \theta_{23} = 0.5$ are almost identical to those obtained in 
\cite{T2KK}. 
It is evident that the sensitivities do not depend strongly on 
$\sin ^2 \theta_{23}$ as far as the value is between 0.40 and 0.60.
In fact, the mass hierarchy can be determined even if 
the $\theta_{23}$ octant degeneracy is not resolved. 
But, the sensitivity to mass hierarchy resolution gradually improves 
as $\sin ^2 \theta_{23}$ becomes larger, as seen in 
Fig.~\ref{sensitivity-mass-hierarchy}. 
It is natural because $\Delta P^{\,\text{norm~inv}}$ in (\ref{DeltaP_sign}), 
or the appearance probability itself is proportional to $\sin ^2 \theta_{23}$. 
An alternative way of presenting the same result is to use 
$s^2_{23} \sin ^2 2\theta_{13}$ for the ordinate. 
An approximate scaling behavior is observed as expected 
by $\Delta P^{\,\text{norm~inv}}$ in (\ref{DeltaP_sign}).

\section{Summary and Discussion}
\label{summary}

In this paper, we have shown that a setting with two identical water 
Cherenkov detectors of 0.27 Mton fiducial mass, one in Kamioka 
and the other in Korea, which receive almost the same neutrino beam 
from J-PARC has capability of resolving the $\theta_{23}$ 
octant degeneracy {\em in situ} by observing difference of the 
solar oscillation term between both detectors. 
The feature of the sensitivity region indicates that 
the present method is quite complementary to the reactor-accelerator 
combined method explored in \cite{resolve23}. 
Together with the potential for resolution of the intrinsic and the 
sign-$\Delta m^2_{31}$ degeneracies previously reported in \cite{T2KK} 
(with confirmation in Sec.~\ref{revisit} by an improved treatment),  
we have demonstrated that the Kamioka-Korea two detector complex 
can resolve all the eight-fold neutrino parameter degeneracy 
under the assumption that $\theta_{13}$ is within reach by the next 
generation accelerator experiments and $\theta_{23}$ 
is not too close to $\pi/4$.

As an outcome of these studies, the strategy toward determination 
of the remaining unknowns in the lepton flavor mixing can be
discussed. 
It is nice to see that such program can be defined only with the single 
experiment based on the conventional superbeam 
technology which does not require long-term R$\&$D efforts,  
and the well established detector technology.
It opens the possibility of accurate determination of the neutrino mixing 
parameters, $\theta_{23}$, $\theta_{13}$, $\delta$, 
as well as the neutrino mass hierarchy, 
by lifting all the eight-fold degeneracy  
which should merit our understanding of physics of lepton sector.

Our treatment in this paper includes a new systematic error which 
accounts for possible difference in spectral shape of the neutrino beam 
received by the two detectors in Kamioka and in Korea.  
We have shown that, despite the existence of such new uncertainty 
which might hurt the principle of near-far cancellation of the 
systematic errors, the capability of determining neutrino mass hierarchy 
and sensitivity to CP violation are kept intact.

We have also reported a progress in understanding the theoretical 
aspect of the problem of how to solve the parameter degeneracy. 
Because of the property phrased as ``decoupling between degeneracies'' 
which is shown to hold in a setting that allows perturbative treatment of 
matter effect, 
one can try to solve a particular degeneracy without worrying 
about the presence of other degeneracies. 
This feature may be contrasted to those 
of the very long baseline approaches, such as the neutrino factory, 
in which one would not expect the discussion in this paper to 
hold.

An alternative but closely related approach toward determination 
of the global structure of lepton flavor mixing in a single experiment 
is to utilize an on-axis wide band neutrino beam to explore the 
multiple oscillation maxima, which may be called the ``BNL strategy'' 
\cite{BNL,FNALversion}. 
This strategy can be applied to the far detector in Korea, as examined 
by several authors \cite{Hagiwara,Dufour-Seoul2006,Rubbia-Seoul2006}.\footnote{
Very roughly speaking ignoring the issue of backgrounds
and assuming the same baseline length, 
one would expect that wide band beam option is better in sensitivity 
to the neutrino mass hierarchy, 
while the same off-axis angle option studied in this paper is 
advantageous to resolve the $\theta_{23}$ octant degeneracy 
for which low energy bins are essential.
}
In this case, however, one needs to understand the energy 
dependence of the background and the signal efficiency 
as well as the neutrino interaction cross section precisely 
for both the intermediate and the far detectors.
In particular, since the low energy bins are enriched with 
neutral current background contamination that comes from 
events with higher neutrino energies 
\cite{Dufour-Seoul2006} 
the cancellation of the systematic errors between 
the two detectors, which is the key ingredient in our analysis, 
does not hold. 
Nonetheless, we emphasize that the potentially powerful method is 
worth to examine further with realistic estimate of the 
detector performance.

Finally, we remark that the J-PARC 2.5 degree off-axis beam
with the baseline length of 1,000 to 1,250~km should be 
available in the Korean Peninsula. Therefore, it may be possible to
further enhance the sensitivity to the $\theta_{23}$ octant
by taking a longer baseline length for the Korean detector.
The best baseline length and the detector location 
should be decided so that the experiment has the best 
sensitivities to the oscillation parameters, especially
to the CP phase $\delta$, mass hierarchy and the octant of 
$\theta_{23}$.

\begin{acknowledgments}
We would like to thank M.~Ishitsuka and K.~Okumura for 
the assistance in the analysis code. 
H.N. thanks Stephen Parke and Olga Mena for useful discussion. 
This work was supported in part by the Grant-in-Aid for Scientific Research, 
Nos. 15204016 and 16340078, Japan Society for the Promotion of Science, 
Funda\c{c}\~ao de Amparo \`a Pesquisa do Estado de Rio de Janeiro (FAPERJ) 
and by Conselho Nacional de Ci\^encia e  Tecnologia (CNPq). 
\end{acknowledgments}

\end{document}